\documentclass[
preprint,
superscriptaddress,
showpacs,preprintnumbers,
 amsmath,amssymb,
 aps,
 prl,
]{revtex4-1}

\usepackage{graphicx}
\usepackage{dcolumn}
\usepackage{bm}
\usepackage{hyperref}
\usepackage[usenames, dvipsnames]{color}

\begin{document}

\title{Evidence for unconventional superconductivity in Half-Heusler YPdBi and TbPdBi compounds revealed by London penetration depth measurements}
\author{S. M. A. Radmanesh}
\affiliation{Advanced Materials Research Institute and Department of Physics, University of New Orleans, New Orleans, Louisiana 70148, USA}
\author{C. Martin}
\affiliation{Ramapo College of New Jersey - Mahwah, New Jersey, 07430, USA}
\author{Yanglin Zhu}
\affiliation{Department of Physics and Engineering Physics, Tulane University, New Orleans, Louisiana 70118, USA}
\author{X. Yin}
\affiliation{Center for High Pressure Science $\&$ Technology Advanced Research (HPSTAR), Beijing 100094, China}
\author{H. Xiao}
\affiliation{Center for High Pressure Science $\&$ Technology Advanced Research (HPSTAR), Beijing 100094, China}
\author{Z. Q. Mao}
\affiliation{Department of Physics and Engineering Physics, Tulane University, New Orleans, Louisiana 70118, USA}
\affiliation{Department of Physics, Pennsylvania State University, PA 16802, USA}
\author{L. Spinu}
\affiliation{Advanced Materials Research Institute and Department of Physics, University of New Orleans, New Orleans, Louisiana 70148, USA}

\date{\today}

\begin{abstract}
The half-Heusler compounds YPdBi and TbPdBi, while having similar band structure, exhibit different magnetic properties. YPdBi is a diamagnet, while TbPdBi shows antiferromagnetic order below 5.5 K. Both are superconductors with T${_c}\approx$1 K for YPdBi and T${_c}\approx$1.75 K for TbPdBi. Such a contrast in properties between these two compounds opens a question about the effects of band structure or magnetic correlations on superconductivity.  Using the combination of a tunnel diode oscillator (TDO) and a commercial  dilution refrigerator, we measured temperature dependent magnetic penetration depth $\Delta\lambda(T)$ in single crystals of YPdBi and TbPdBi, down to temperatures  as low as 0.1K. We found that penetration depths of both compounds do not show exponential temperature dependence and saturation at low temperatures, as expected for conventional BCS superconductors. Instead, in both compounds, the penetration depth can be described by a power  law $\Delta\lambda(T) = A\times T^{n}$. The coefficient A was found to be about 50$\%$ smaller in TbPdBi, but the exponents are very similar, $n = 2.76\pm 0.04$ in YPdBi and $n = 2.6\pm 0.3$ in TbPdBi, respectively. Our results suggest unconventional superconductivity  in both YPdBi and TbPdBi.

\end{abstract}
\maketitle
In addition to their previously explored interesting physical properties, such as giant, linear magnetoresistance or heavy fermion carriers, the half-Heusler compounds were predicted to have topological insulating states~\cite{Chadov10, Lin10}. Furthermore, superconductivity has also been discovered in the $R$PdBi and $R$PtBi ($R$ = rare earth) half-Heusler materials~\cite{Butch11, Bay12, Nakajima15}. Strong spin-orbit coupling combined with the absence of inversion symmetry can give rise to unconventional superconductivity, and one of the most unique features is the possibility of superconducting pairing with the total angular momentum other than singlet ($J$=0) or triplet ($J$=1), which is in general allowed by pairing of $j$=1/2 fermions.

Through substitution of the rare earth $R$ the band inversion $\Delta E = \Gamma_{8} - \Gamma_{6}$, between the energies of the s-like ($j$=1/2) $\Gamma_6$ and the p-like ($j$=3/2) valence bands of Bi can be tuned in $R$PdBi and $R$PtBi from negative (trivial) to positive (nontrivial). For a positive band inversion, when the chemical potential resides in the $\Gamma_{8}$ band, the conduction carriers have an angular momentum $j$=3/2. It is in particular this instance that allows for pairing with total angular momentum $J=j_{1}+j_{2}>1$. Recent measurements of temperature dependent penetration depth ($\Delta\lambda(T)$) on the nontrivial YPtBi half-Heusler compound found a linear behavior, consistent with line nodes in the superconducting gap~\cite{Kim18}. According to Ref.~\cite{Kim18}, the nodal gap is most likely indicative of pairing with higher total angular momentum. The symmetry and the structure of the superconducting gap is far from being resolved in the $R$PdBi and $R$PtBi materials, and one way to gain insight is to investigate the superconductive gap for the case of the trivial band inversion.

Another related open question regarding the $R$PdBi and $R$PtBi half-Heusler superconductors is the pairing mechanism. By changing the rare earth $R$, the magnetic properties can also be tuned, in addition to the effects on the band structure. Recent experimental study has found that in RPdBi (R= Y, Sm, Gd, Tb, Dy, Ho, Er, Tm, and Lu), with increasing antiferromagnetic (AF) coupling, superconductivity is suppressed~\cite{Nakajima15}. Therefore, there seems to be a competition between the AF and superconducting ground states. Similar work however, has shown evidence of superconductivity even in a compound with AF ordering and higher N\'{e}el temperature, suggesting possible coexistence of the two states~\cite{Xiao18}. It is important in such systems to investigate the potential role played by magnetic correlations on superconductivity.

In order to address the open questions above, we present here temperature dependent penetration depth ($\Delta\lambda(T)$) studies on single crystals of two $R$PdBi half-Heusler superconductors with trivial band inversion: YPdBi, which is non-magnetic, and TbPdBi, which orders antiferromagnetically below the N\'{e}el temperature $T_{N}$ = 5.5 K and becomes superconducting at a critical temperature as high as $T_{c}\approx$ 1.75 K. Through the study of these two compounds we can compare the behavior of penetration depth between these two systems with different levels of magnetic correlations, and furthermore, with previous findings on the nontrivial band inversion.

Single crystals of YPdBi and TbPdBi were grown using Bi as flux. One sample of TbPdBi, with dimensions 0.281$\times$0.156$\times$0.219 mm$^3$ was selected from the same crystal growth described in Ref.~\cite{Xiao18}, where X-rays diffraction and magnetic field dependent transport confirmed the crystal structure, stoichiometry and the superconducting transition. Penetration depth of YPdBi was measured on three single crystals with following dimensions: 0.250$\times$0.188$\times$0.125 mm$^3$ (Sample 1), 0.188$\times$0.156$\times$0.125 mm$^3$ (Sample 2), and 0.125$\times$0.125$\times$0.156 mm$^3$ (Sample 3) respectively. In order to verify the superconducting transition of YPdBi, we measured separately resistance of two additional samples selected from the same batch, with dimensions 0.180$\times$0.126$\times$0.040 mm$^3$ (Sample 4) and 0.410$\times$0.370$\times$0.059 mm$^3$ (Sample 5). Temperature dependent penetration depth measurements were performed by placing the samples inside the inductor of an LC-tank resonator, biased by a tunnel diode, i.e. the tunnel diode oscillator (TDO) set-up~\cite{Degrift75}. The TDO was mounted on a Janis Model JDry-500 cryogen-free He3 - He4 dilution refrigerator system. A special set-up, where the sample is thermally linked to the mixing chamber and the TDO circuit is connected to the second (4 K) cooling stage of the dilution refrigerator, was employed~\cite{Diaconu}. This way, the circuit temperature is stabilized within better than 0.2 mK at 4 K, while the sample temperature is varied over several degrees. The resonant frequency of the empty resonator is 5.9 MHz, with a noise floor of about 0.5 Hz and no significant drifts over the duration of the experiment. The base temperature in our experiments was 70 mK.

The change of the TDO frequency is directly related to the change in penetration depth: $\Delta f(T) = (-G/R)\Delta\lambda(T)$. The parameter $G$ is a calibration constant that depends on the volume of the sample relative to the volume of the inductor (filling factor), and on the shape of the sample (demagnetization factor). More precisely, this calibration constant represents the difference between the frequencies of the empty resonator and that with the sample inside the solenoid, that was determined directly by cooling the system both with and without sample. The parameter $R$ represents an effective dimension calculated from the actual dimensions of the sample according to Ref.~\cite{Prozorov06}.
\begin{figure}[htp]
\includegraphics[width=0.45\textwidth]{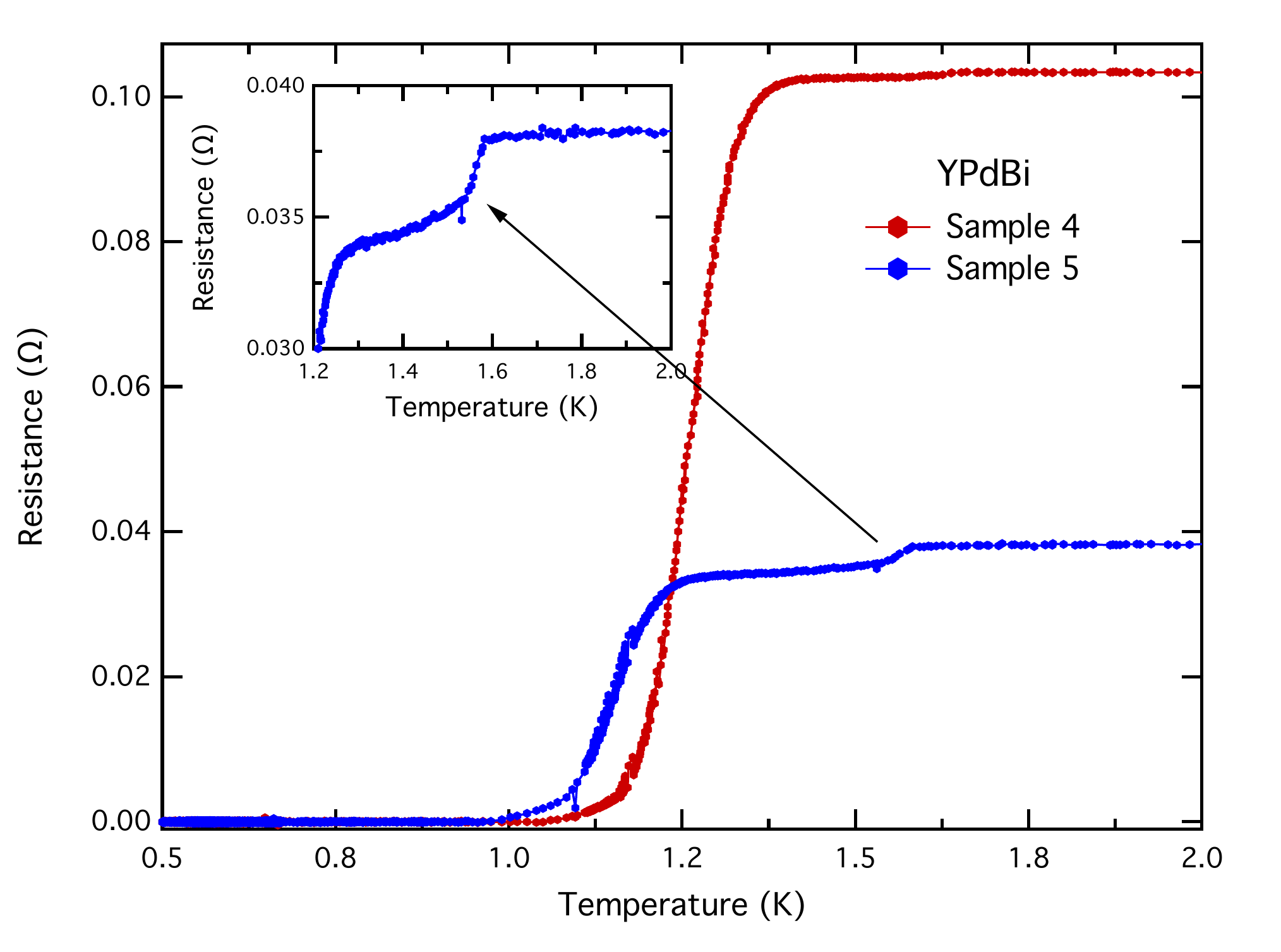}
 \caption{(Color online) Main panel: Low temperature resistance showing the superconducting transition of two samples of YPdBi (Sample 4 and Sample 5). Inset: A zoom on the data for Sample 5 around 1.6 K, showing the small drop in resistance associated with presence of an impurity phase, as discussed in the main text.}
\label{Fig1}
\end{figure}

\begin{figure}[htp]
\includegraphics[width=0.45\textwidth]{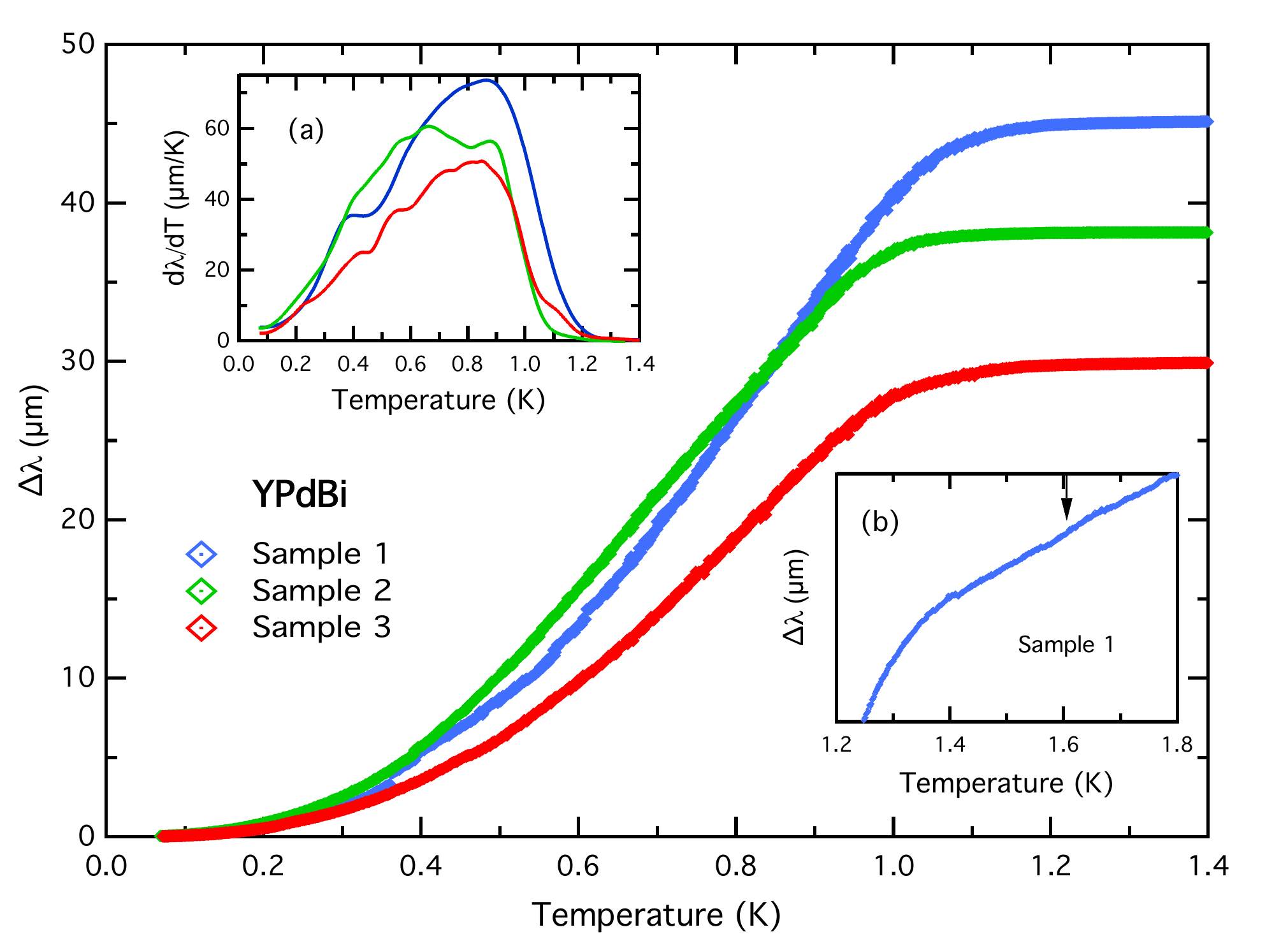}
 \caption{(Color online) Main panel: Temperature dependence of penetration depth, between 0.07 K and T$_c$, for three samples of YPdBi. Inset (a): The derivative of $\Delta\lambda(T)$ showing the width of the superconducting transition. Inset (b): The data for Sample 1 from the main panel at higher temperature, showing no sizable diamagnetic screening at 1.6 K, marked with an arrow.}
\label{Fig2}
\end{figure}

Figure 1 shows the temperature dependence of resistance for two samples of YPdBi (Sample 4 and Sample 5). Clear superconducting transitions, but with slightly different critical temperatures, can be observed in both samples. The onset critical temperature is $T_{c}\approx1.3$ K in Sample 4 and $1.2$ K in Sample 5. The complete loss of resistance occurs at 1.04 K and 0.97 K, respectively. Therefore, we can estimate a difference of about 0.07 to 0.1 K between their critical temperatures. The transition widths are very similar, averaging at 0.25$\pm$0.02 K. Another important feature in Fig.~\ref{Fig1} is the small, sharp drop of resistance at $T\approx1.6$ K. This was observed in both samples, although more strongly in Sample 5, displayed in the inset of Fig.~\ref{Fig1}. The temperature matches closely the critical temperature of the $\alpha-Bi_{2}Pd$ phase, therefore this feature is indicative for the presence of a small amount of the impurity phase $\alpha-Bi_{2}Pd$, which was created in baking the contacts as previously reported in Ref.~\cite{Nakajima15}. We notice that the drop in resistance is about 0.7$\%$ in Sample 4 and 8$\%$ in Sample 5, suggesting a small amount of impurity concentration. Moreover, we checked one of our samples selected for the penetration depth study discussed below and did not observe a measurable diamagnetic screening at 1.6 K (see inset (b) of Fig.~\ref{Fig2}), ruling out the effect of the impurity $\alpha-Bi_{2}Pd$ phase on our study of YPdBi.

Figure 2 shows the magnetic field penetration depth for three samples of YPdBi, measured between 0.07 K and 1.4 K. As discussed above, to address the concerns related to the possible presence of the impurity phase, Sample 1 has been measured to higher temperature, and we display in the inset (b) a zoom on the region around 1.6 K, marked with an arrow. Any possible feature at that temperature is within the noise level of our data, and is definitely negligible comparing to the clear diamagnetic screening, with the onset around 1K, that can be observed in the main panel for all three samples. The relatively broad superconducting transitions make it difficult to precisely establish the critical temperature $T_c$ from penetration depth data, which is otherwise relevant for our discussion in the next paragraph. For a better estimate of $T_c$ and a comparison between samples, we calculate the rate of change of penetration depth with temperature, $d\lambda/dT$, shown in the inset (a) of Fig.~\ref{Fig2}. The onset of superconductivity, defined as the point where $d\lambda/dT$ starts deviating from the normal state value upon cooling, varies only slightly between samples, from 1.21 K to 1.10 K. The position of the peak of $d\lambda/dT$, considered to be the highest temperature at which the sample is still fully superconducting would give a critical temperature between 0.86 K and 0.70 K. Therefore, the transition width is around 0.35-0.40 K. As expected, this is larger than the width measured from resistance data. Once the supercurrent finds a continuous path to flow through the sample, the measured resistance will drop to zero, whereas penetration depth will still vary with temperature due to the temperature dependence of the superconducting gap, anisotropy of the gap, or other factors, such as impurity scattering. Furthermore, the strong temperature dependence of penetration depth, which is discussed in detail below, makes the transitions shown in the main panel of Fig.~\ref{Fig2} appear even more broad.
\begin{figure}[htp]
\includegraphics[width=0.45\textwidth]{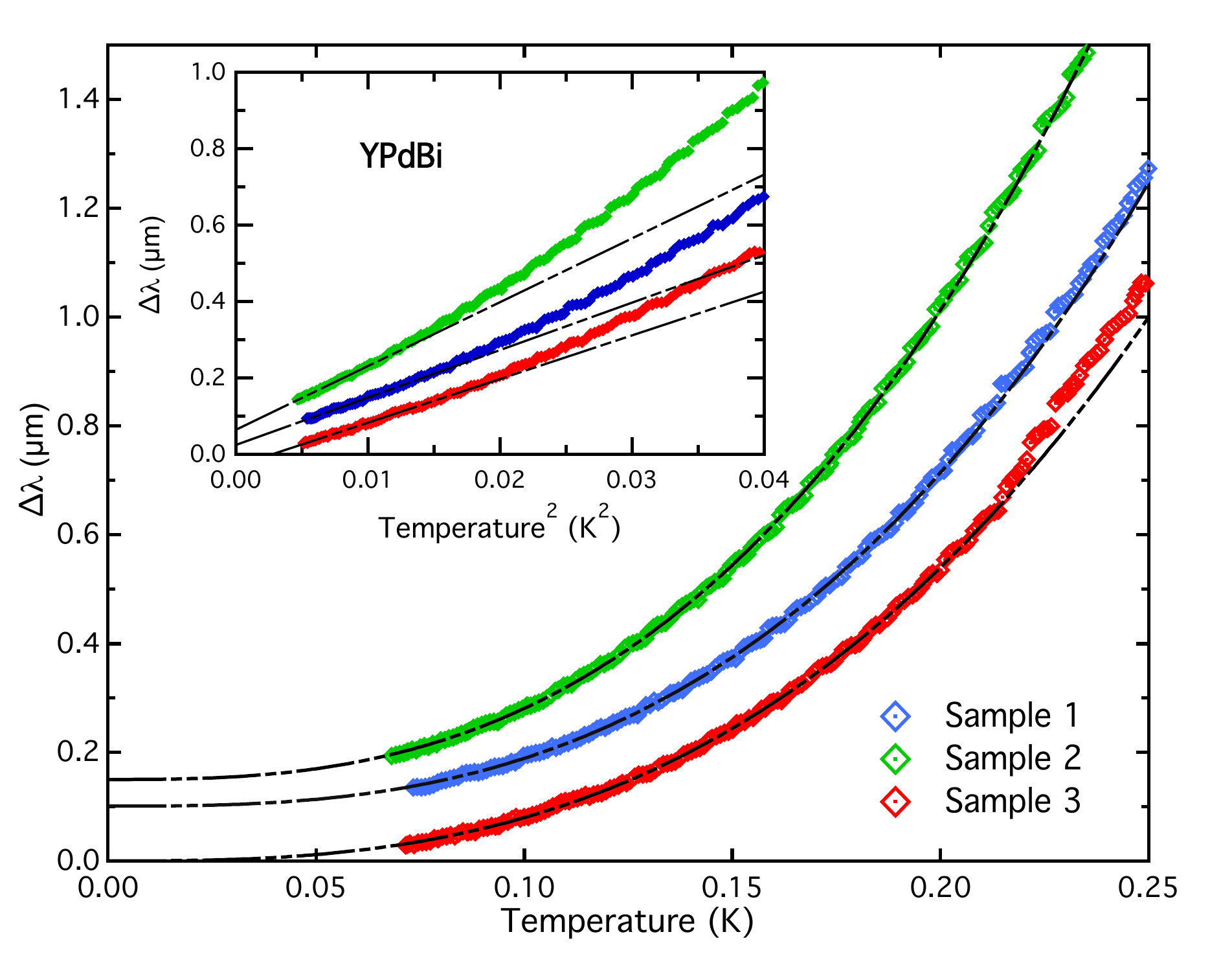}
\caption{Main Panel: Low temperature measurements of $\Delta\lambda(T)$ for three YPdBi samples (symbols), and power law fits of the data (discontinuous lines), as discussed in the main text. The traces were shifted vertically for clarity. Inset: Low temperature $\Delta\lambda$ versus T$^2$ (symbols) and linear fits (discontinuous lines).}
\label{Fig3}
\end{figure}

For a conventional superconductor, the superconducting gap is nearly temperature independent below $\approx$~0.3~$T_c$. In consequence, penetration depth $\lambda(T)$ is expected to display very small temperature dependence, and to saturate exponentially for $T\leq 0.3T_c$. Based on the discussion of the critical temperature and the transition width above, we conservatively focus on the behavior of $\Delta\lambda(T)$ in YPdBi between base temperature (0.07 K) and 0.20 K, as displayed in Fig.~\ref{Fig3}. First, it must be noticed that there is significant temperature dependence, and while $\Delta\lambda$ increases in the range up to 0.20 K by about 525 nm in sample 3 and 620 nm in sample 1, the increase in sample 2 is larger than 800 nm.  As it has been pointed out in a previous study on YPtBi, the large temperature dependence of penetration depth may be explained in general by the low carrier concentration in these materials ($n\approx10^{18}-10^{19} cm^{-3}$), and hence the large London penetration depth $\lambda_{L}=\lambda(0)=\sqrt{m/(\mu_{0}ne^{2})}$~\cite{Kim18}. Then, possible difference in carrier concentration between samples from the same compound may explain the large variation observed in our samples.

It can be clearly observed in Fig.~\ref{Fig3} that penetration depth does not saturate toward zero temperature, and in consequence, the BCS exponential function did not correctly reproduce its low temperature behavior. Instead, we found that a power-law fit $\Delta\lambda(T) = A\times T^{n}$ is more accurate. The exponent was very similar for all three samples, $n$ = 2.76$\pm$0.04, despite the fact that the factor $A$ was found to be more than 50$\%$ larger in sample 2 than in the other two.

\begin{figure}[htp]
\includegraphics[width=0.45\textwidth]{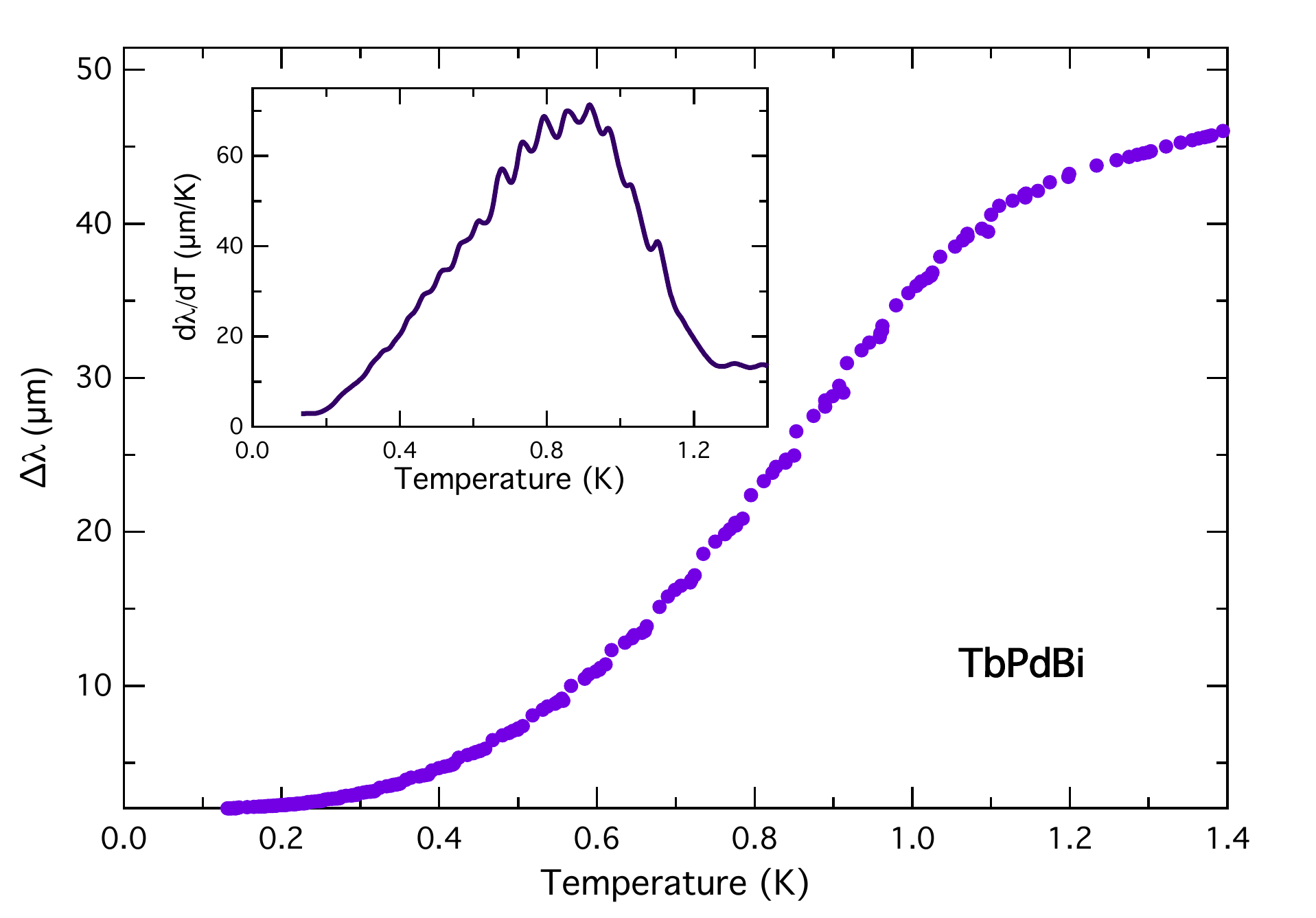}
\caption{Main panel: $\Delta\lambda(T)$ for TbPdBi between 0.131 K and T$_c$. Inset: $d\lambda/dT(T)$ showing the width of the superconducting transition.}
\label{Fig4}
\end{figure}

Before discussing the potential implications of our findings on YPdBi, it is important to also look at the low temperature penetration depth of a similar ternary half-Heusler compound TbPdBi. Figure~\ref{Fig4} shows $\Delta\lambda(T)$ (main panel) and its derivative $d\lambda/dT(T)$ (inset), from base temperature to 1.4 K. Because of an experimental issue the base temperature for this experiment was 0.131 K. Similar to our discussion on YPdBi, we find that superconductivity in TbPdBi emerges around 1.25 K and the sample is fully superconducting below 0.85 K (position of the peak of $d\lambda/dT(T)$). Therefore, both the critical temperature and the superconducting transition width are very similar between the two compounds. Also, our base temperature of 0.131 K is still well below 0.3 T$_c$ in TbPdBi, even by most conservative estimate of T$_c$.

\begin{figure}[htp]
\includegraphics[width=0.45\textwidth]{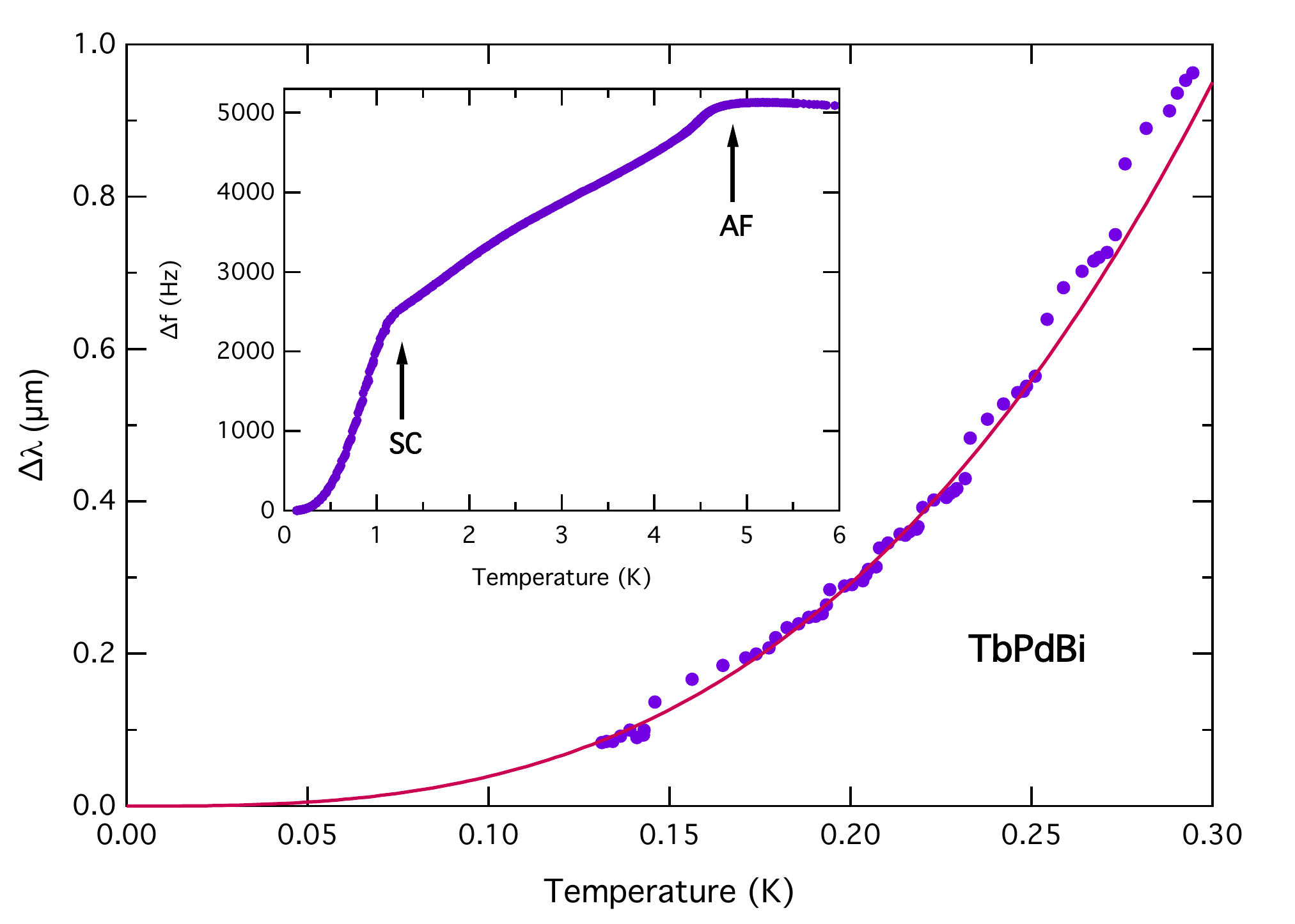}
\caption{Main panel: Low temperature measurements of $\Delta\lambda(T)$ for a TbPdBi sample (symbols), and power law fits of the data (red line), as discussed in the main text. Inset: Change in TDO frequency between base temperature and 6 K when the TbPdBi sample is loaded. The superconducting (SC) and antiferromagnetic (AF) transitions are shown with arrows.}
\label{Fig5}
\end{figure}

As it can be seen from the main panel of Fig.~\ref{Fig5}, $\Delta\lambda(T)$ in TbPdBi also shows significant temperature dependence at low temperature, but the rate of change is smaller than in YPdBi. We found that up to 0.2 K, it changes by less than 300 nm, which is less than 50$\%$ of the values for YPdBi in Fig.~\ref{Fig3}. As discussed above, precise determination of zero-temperature penetration depth would help explain the difference. Just like for YPdBi, a power-law fit describes well our experimental data from Fig.~\ref{Fig5}. The exponent was found to be $n$ = 2.6$\pm$0.3, showing a larger variation with the temperature range of the fit. Nevertheless, the average value of the exponent is similar to that determined above for YPdBi.
Based on the data shown in Figs.~\ref{Fig2} - \ref{Fig5} we can conclude that both YPdBi and TbPdBi have similar critical temperatures and similar temperature dependence of penetration depth. We believe that these findings have implications on the potential role played by magnetic correlations in mediating superconductivity of the half-Heusler materials RPdBi  (R=rare earth element). As we mentioned in the introduction, based on Ref.~\cite{Nakajima15} YPdBi is considered to have the weakest magnetic correlations and the highest superconducting transition temperature. On the other hand, the magnetic correlations in TbPdBi are strong enough to result in AF ordering below 5.5 K, and, possibly to suppress superconductivity. Similar work however, has reported bulk superconductivity in TbPdBi, with T$_c$ up to 1.75 K, despite magnetic ordering~\cite{Xiao18}. The inset of Fig.~\ref{Fig5} displays the change in resonant frequency of our TDO circuit from base temperature up to 6 K when the TbPdBi sample is loaded, showing clearly both the superconducting (SC) and the antiferromagnetic (AF) transitions. We note that similar data on YPdBi did not reveal any additional magnetic transition, other than superconductivity, up to 10 K. Based on the discussion in Ref.~\cite{Xiao18}, despite the increase in magnetic correlations from YPdBi to TbPdBi, superconductivity remains robust in both systems, which is further confirmed by our present data. Moreover, we found that penetration depth also has similar power law temperature dependence in both compounds. Therefore, we suggest indirectly that AF fluctuations do not play a leading role on superconducting pairing mechanism, and further more direct probes, such as neutron scattering may elucidate this question. The relatively large variation between the rate of change of penetration depth with temperature, i.e the parameter $A$, can rather be explained by different values of the London penetration depth. One would expect that other factors that can affect $A$, such as impurity scattering, variation of the superconducting gap or the gap anisotropy with size of the Fermi surface should also affect the exponent $n$, contrary to our finding.

Furthermore, we suggest that the power law behavior of $\Delta\lambda(T)$ with very similar exponents observed in two compounds with different scale of AF fluctuations, as well as in different samples from the same compound, is intrinsically related to the symmetry of the superconducting gap, which is unconventional, rather than to other effects such as of impurity scattering, like it was found for example in the Fe-based superconductors. It is worth noticing otherwise that in some Fe-based superconductors, where the role of magnetic correlations on superconductivity were clearly established, penetration depth was also found to have a power law temperature dependence. The exponent however varied significantly between samples with different doping and/or impurity levels. The impurity scattering between bands with sign changing superconducting gap was one of the factors often invoked to explain the behavior of $\Delta\lambda(T)$~\cite{Hirschfeld11}. Unlike Fe-based superconductors, YPdBi and TbPdBi are single band superconductors, ruling out a similar explanation. Intra-band scattering may rather play an important role, but if that were the case, one would have expected a larger variation of the exponent between our three samples of YPdBi. Most likely, the power law temperature dependence of penetration depth in our samples is determined by the symmetry of the gap, and the lack of saturation and non-exponential temperature dependence rule out a conventional, isotropic s-wave gap.

It is also important to notice that the exponent $n >$ 2 observed in all our samples is very different from the nearly linear temperature dependence of penetration depth previously reported in YPtBi~\cite{Kim18}. A linear behavior of $\Delta\lambda(T)$ implies the existence of line nodes in the superconducting gap, and Ref.~\cite{Kim18} suggests that a nodal order parameter in the half-Heusler compounds represents evidence for unconventional Cooper-pairs with high angular momentum. Strong impurity scattering in a superconductor with line nodes can create low energy excitations and change the linear T-dependence into a quadratic T$^2$, below a characteristic temperature T$^*$~\cite{Hirschfeld93}. In order to verify such hypothesis, we plot in the inset of Fig.~\ref{Fig3} $\Delta\lambda$ versus T$^2$. In the limit T$\rightarrow$0 penetration depth does follow a quadratic temperature dependence, but in a narrow range. Moreover, the linear dependence is expected to recover above T$^*$, whereas our data from the inset of Fig.~\ref{Fig3} shows stronger temperature variation than quadratic or linear at higher temperature. In fact, the main panel of Fig.~\ref{Fig3} and Fig.~\ref{Fig5} clearly shows that an exponent quite larger than two fits our data over a relatively broad temperature range, rather suggesting that nodes in the superconductive gap are unlikely, at least in YPdBi and TbPdBi.

Therefore, the nodal order parameter is not a universal feature of the half-Heusler superconductors. To better understand the difference between our findings and those from Ref.~\cite{Kim18}, we look at the significant differences between YPdBi, TbPdBi and YPtBi. Hall effect and Shubnikov-de Haas oscillations measurements have revealed very similar carrier concentrations and Fermi surface size between YPdBi and YPtBi~\cite{Kim18, Wang13}. There is however a major difference between the band structures of these two compounds. In YPtBi, strong spin-orbit scattering gives rise to a relatively large and positive (nontrivial) band inversion at the $\Gamma$ point, $\Delta E = \Gamma_{8} - \Gamma_{6} > 0$. Both SdH oscillations~\cite{Kim18} and angle-resolved photoemission (ARPES)~\cite{Liu16} data found that the chemical potential is situated in the p-like $\Gamma_{8}$ band, therefore the total angular momentum of conduction electrons is j=3/2. In consequence, as explained in Ref.~\cite{Kim18}, pairing with total angular momentum beyond singlet s-wave (J=0) and triplet p-wave (J=1) is possible. It was suggested that the larger angular momentum pairing channel in particular might be responsible for a superconducting gap with line nodes~\cite{Schnyder15}. On contrary, the band structure of YPdBi (and TbPdBi) was found to be topologically trivial~\cite{Chadov10}. The s-like band $\Gamma_{6}$ in the valence band is situated above p-like $\Gamma_{8}$ ($\Delta E < 0$), and is closer to the Fermi level. Therefore, the conduction band in YPdBi has an s-symmetry, allowing only singlet or triplet superconducting pairing. In this case, anisotropy in the superconducting gap, as suggested by the power-law dependence of the penetration depth, is more likely to be favored by the p-wave (triplet) pairing channel~\cite{Mineev99}. Therefore, although penetration depth is not directly spin-dependent measurement, our data supports the existence of triplet pairing, most likely mixed with singlet, superconductivity in YPdBi and TbPdBi.

Aside from anisotropy of the superconducting gap, there is yet another possible reason for a power-law behavior of $\Delta\lambda(T)$, when mixed singlet-triplet pairing is considered. YPdBi and TbPdBi are single band conductors, therefore within the same band, Fermi surface of the condensate consists of regions of singlet and triplet. Intra-band impurity scattering between regions with different pairing symmetries is expected to manifest as pair-breaker, similar to the effect of inter-band impurity scattering in multiband superconductors with sign changing gap(s±)~\cite{Preosti96}. As a consequence, it can increase the number of quasiparticles with increasing temperature, resulting in a stronger temperature dependence of penetration depth then otherwise exponential. However, intra-band impurity scattering is expected in this case to also affect the critical temperature T$_{c}$. We reemphasize that our three different YPdBi samples and one TbPdBi, they all had very similar T$_{c}$ and very similar exponent of $\Delta\lambda(T)$, apparently undercutting the role of intra-band scattering. It is again possible that because the region of triplet pairing on the Fermi surface is very small comparing to the singlet one, the effect of scattering between the two regions when summing the gap over the entire FS to be too small in order to affect the critical temperature. While is more likely that the anisotropy of the superconducting gap is responsible for our observation of a power-law temperature dependence of penetration depth in YPdBi and TbPdBi, we believe that future theoretical and experimental work on the role of impurity scattering is also worth pursuing. Nevertheless, our results provide strong support for an unconventional, possibly mixed singlet-triplet superconductivity in the half-Heusler semiconductors with trivial band inversion.

\begin{acknowledgments}
The work at Tulane was supported by the US National Science Foundation under grant DMR - 1707502. LS thanks National Science Foundation for the support through the Independent Research/Development (IR/D) Program. We would like to thank Peter Hirschfeld for stimulating discussions and suggestions.
\end{acknowledgments}


\end{document}